Linking Homeostasis to Reinforcement Learning: Internal State Control of Motivated Behavior

Naoto Yoshida(1), Henning Sprekeler(2, 3, 4, 5), Boris Gutkin(6)

(1) Kyoto University, Yoshida-honmachi, Sakyo-ku, Kyoto, 606-8501, Japan

(2) Technische Universität Berlin, Marchstr. 23, 10587 Berlin, Germany

(3) Theoretical Sciences Visiting Program, Okinawa Institute of Science and Technology Graduate University, Onna, 904-0495, Japan

(4) Bernstein Center for Computational Neuroscience Berlin, Philippstr. 13, 10115 Berlin, Germany

(5) Science of Intelligence, Research Cluster of Excellence, Marchstr. 23, 10587 Berlin, Germany

(6) Group for Neural Theory, Laboratoire des Neurosciences Cognitives et Computationnelles (LNC2), INSERM U960, Ecole Normale Superieure PSL* University, Paris, France

email:yoshida.naoto.8x@kyoto-u.ac.jp, boris.gutkin@ens.psl.eu

**Abstract:**

For living beings, survival depends on effective regulation of internal physiological states through motivated behaviors. In this perspective we propose that Homeostatically Regulated Reinforcement Learning (HRRL) as a framework to describe biological agents that optimize internal states via learned predictive control strategies, integrating biological principles with computational learning. We show that HRRL inherently produces multiple behaviors such as risk aversion, anticipatory regulation, and adaptive movement, aligning with observed biological phenomena. Its extension to deep reinforcement learning enables autonomous exploration, hierarchical behavior, and potential real-world robotic applications. We argue further that HRRL offers a biologically plausible foundation for understanding motivation, learning, and decision-making, with broad implications for artificial intelligence, neuroscience, and understanding the causes of psychiatric disorders, ultimately advancing our understanding of adaptive behavior in complex environments.

**Highlights:**

- The Homeostatically Regulated Reinforcement Learning (HRRL) gives a framework that integrates homeostasis with reinforcement learning to explain how organisms learn behaviors

- to maintain internal stability and assure survival.

- Based on HRRL we argue for an interoceptive definition of reward based on reducing drives that are defined as deviation from homeostatic optimum.

- HRRL can account normatively for a number of apparent deviations from rationally directly from properties of the drive.

- HRRL provides a framework to (1) link interoceptive-based learning to future development of AI and autonomous robots, (2) provide an interoceptive homeostatic view of psychological disorders, (3) develop a theory for interactions between internal states and high-level cognitive processes.

**Introduction**

The primary goal of all living beings is to survive and maintain fitness. Organisms actively maintain their ability to pursue these aims through behaviors that stabilize their internal states in relation to the external world. This drive for stability, termed homeostasis, regulates essential physiological variables (such as hunger, core temperature, sodium levels e.g., see Palmiter [1], Tan and Knight [2], Bernal et al. [3], Viskaitis et al. [4]) within narrow bounds, ensuring survival through a complex network of control processes that span from molecular to behavioral responses.

This regulatory challenge is compounded by the complexity of natural environments, where resources are often scarce and conditions uncertain. We can operationally separate the external environment from the internal environment of the organism, which can be defined as a compendium of its internal state. Organisms must estimate internal states with sufficient accuracy (through interoception, Craig [5]), adapt to multiple interacting timescales, and handle varying degrees of synergy or opposition among internal variables.

Classical homeostatic theories, formalized based on control theoretic models – as reactive or predictive control – can yield behavioral patterns and foraging strategies that are optimal for given environments, but how animals learn such behaviors or policies is largely outside the purview of such models (but see Toates [6] for further discussion). Reinforcement learning (RL) theory suggests that optimal behaviors can be explained by agents that are driven (or motivated) to maximize long-term reward accumulation (Sutton and Barto [7]). Reinforcement learning algorithmic models provide a theory for how behaviors are learned from extrinsic rewards. On the other hand, how the rewards are defined is a point that is by and large arbitrarily treated in the traditional RL models.

This paper argues that combining basic reinforcement learning with homeostatic principles enables biological agents to learn complex survival strategies. We review the homeostatically regulated reinforcement learning (HRRL) framework (Keramati and Gutkin [8]), link it to control-theoretic models of motivation, summarize its key properties and applications, and suggest how it might extend to human motivation beyond physiological needs.

**What is reward?**

The major question that is left moot by the standard RL theory is how do we define rewards for biological beings. In other words, how to decide which external resources are beneficial for the agent and worth foraging for and which are detrimental and should be avoided? Which aspects of the environment (and associated choices) should be assigned a positive reward and which a negative one?

In standard RL theory, this is technically known as the reward landscape or schedule and is a priori decided by the designer of the model at hand. On the other hand, homeostatic control theories propose how an organism should act based on its needs to ensure stability in an optimal fashion. Yet even predictive homeostatic control setups do not, by and large, give a systematic account of how a being should learn about what it actually could consume in order to fulfill its needs and/or how to structure behavioral policies (foraging strategies) to navigate the environments in order to gain access to the required resources.

**Homeostatically Regulated RL: Theory overview**

The theory of homeostatically regulated reinforcement learning (Keramati and Gutkin [8,9]; Hulme et al. [10]) provides a link between the two systems by giving a cogent definition of reward and learning optimal behavioral strategies that result in optimal control of internal states (Figure 1a).

At the heart of the HRRL theory is the concept of drive. This has been proposed conceptually in classical theories of motivation as the basis of motivation by Hull [11], as a way to link organismal needs to behaviors that fulfill those needs. In the drive reduction theory, a need is defined as a deviation of the organism's internal state from its homeostatic optimum, or the setting point. In turn, drive is defined as a process that links the need to the motivation that drives the appropriate behavior.

Formally in the HRRL theory, we define a drive function (Keramati and Gutkin [8]). The drive function is defined on an internal state space comprising the relevant internal variables. This function should satisfy several requirements to make sense and also for mathematical expediency. First of all, it should be a function with a minimum at the optimal internal state (the set point). Second, it should be a monotonically increasing function away from this minimum. In sum, the drive function can be thought of as a "motivation" energy function and reflects a distance of the internal state from the optimum. One interpretation of the drive function is that Hulme et al. [10] argued its properties are compatible with ecological survival probability curves.

Having defined a drive function, we can then give a physiologically and biologically cogent definition of a reward. Given an outcome of an action or an event, the reward can be formally defined as the negative change in the drive contingent on that outcome. In other words, consuming a certain amount of a nutrient by a hungry organism (with a negative deviation of the internal state and hence a positive drive towards that state), will move its internal state closer to the homeostatic optimum and decrease the drive. Therefore, the impact of that outcome will be a negative change in the drive and a positive reward. Note that consuming the same amount of nutrient while at the optimum, or beyond it (overfed) will lead to an increase in the drive and hence a negative reward.

---

**Mathematical formulation of the HRRL theory.**
Defining the Homeostatic Reinforcement Learning Framework

1. Drive Function

Let the internal (physiological) state at time *t* be: $H_t = (h_{1,t}, h_{2,t}, ..., h_{n,t})$
And let us define the Homeostatic Setpoint: $H^* = (h_1^*, h_2^*, ..., h_n^*)$ that is optimal for the organism. Then we can define the Drive Function on this internal space as

$$d(H_t) = \left( \sum_{i=1}^{N} |h_i^* - h_{i,t}|^m \right)^{1/n}$$

This function measures the distance from homeostasis. We note that here m and n are free parameters. When *m = n = 2*, it becomes the Euclidean distance.

2. Reward

Given outcome $K_t = (k_{1,t}, k_{2,t}, ..., k_{n,t})$, the physiological state updates as:
$$H_{t+1} = H_t + K_t$$
Reward is defined as the reduction in drive:
$$r(H_t, K_t) = d(H_t) - d(H_{t+1})$$
$$r(H_t, K_t) = d(H_t) - d(H_t + K_t)$$

Given this definition of reward $r(H_t, K_t)$, one can use it in a learning algorithm of choice. E.g. for Q-learning one can define the standard reward prediction error using the homeostatic reward definition and use the standard Q-update (Watkins and Dayan 1992).

3. Equivalence of Reward Maximization and Deviation Minimization

Let $P(H_0)$ be the set of all state trajectories from an initial internal state point $H_0$ to the homeostatic setpoint $H^*$. Define:

$$SDD_p(H_0) = \sum_{t=0}^{n-1} \gamma^t d(H_{t+1})$$

$$SDR_p(H_0) = \sum_{t=0}^{n-1} \gamma^t [d(H_t) - d(H_{t+1})]$$

Then for all $\gamma \in (0,1)$:
$$\arg\min_p SDD_p(H_0) = \arg\max_p SDR_p(H_0)$$

This proposition formally unifies reinforcement learning (RL) and homeostatic regulation (HR): Maximizing reward (defined as drive reduction) is mathematically equivalent to minimizing deviation from physiological setpoints. This offers a normative, biologically plausible model for behavior driven by both internal and external states. We also note that Yoshida et al (2024) showed how the homeostatic reward definition can be related to reward shaping (Ng 1999).

4. Behavioral Properties of HRRL Reward Function

Given the we choose *m>n>1* the shape of the drive function leads to reward and value function that gives several interesting properties to the HRRL agents.

A. Reward Increases with Outcome Magnitude

The reward function is an increasing function of the magnitude of the outcome:
$$\frac{\partial r(H_t, K_t)}{\partial k_{j,t}} > 0 \quad \text{for } k_{j,t} > 0$$

B. Excitatory Effect of Deprivation Level

The reinforcing strength of a constant outcome increases with deprivation level:
$$\frac{\partial r(H_t, K_t)}{\partial |h_j^* - h_{j,t}|} > 0 \quad \text{for } k_{j,t} > 0$$

C. Inhibitory Effect of Irrelevant Drive

Increased deprivation in irrelevant needs can suppress the reinforcing value of an outcome:
$$\frac{\partial r(H_t, K_t)}{\partial |h_i^* - h_{i,t}|} < 0 \quad \text{for all } i \neq j \text{ and } k_{j,t} > 0$$

D. Risk Aversion

The reward function is concave in outcome magnitude, implying risk aversion:
$$\frac{\partial^2 r(H_t, K_t)}{\partial k_{j,t}^2} < 0 \quad \text{for } k_{j,t} > 0$$

-------------------------------------------------------------------------------------------------------------

Having defined the reward, the large set of algorithms in RL become available to acquire behaviors that mediate homeostatic control. Most studies of HRRL so far have focused on value-based RL methods such as temporal difference learning (Sutton and Barto [7]), but other approaches such as policy gradients with actor-critic architectures would be readily applicable. To complete the HRRL setup, we need to define the state and action spaces of the agent, along with their dynamics. Conceptually, it is useful to think of the state space as an outer product of an internal state space – the internal variables that have to be controlled – and an external state space, which describes the situation of the agent in its environment. Each of these state spaces has its individual transition dynamics, but they are coupled via resources in the external state space that can be acquired by the agent to influence its internal states. Note that the wording of "internal" and "external" should be taken with a grain of salt, because the boundary between the agent and its environment is often hard to define (e.g., Palacios et al. [12]).

We note here that such a homeostatic basis of drive can be expanded to an allostatic view, where the need is defined with respect to a dynamically controlled set point. We will see however that combining a homeostatically based drive with RL can produce behaviors that appear to be allostatically driven. Such a situation can arise because RL is not merely an immediate response to inputs but rather a predictive control mechanism over future time series, and because the dynamics of each internal state are interdependently related.

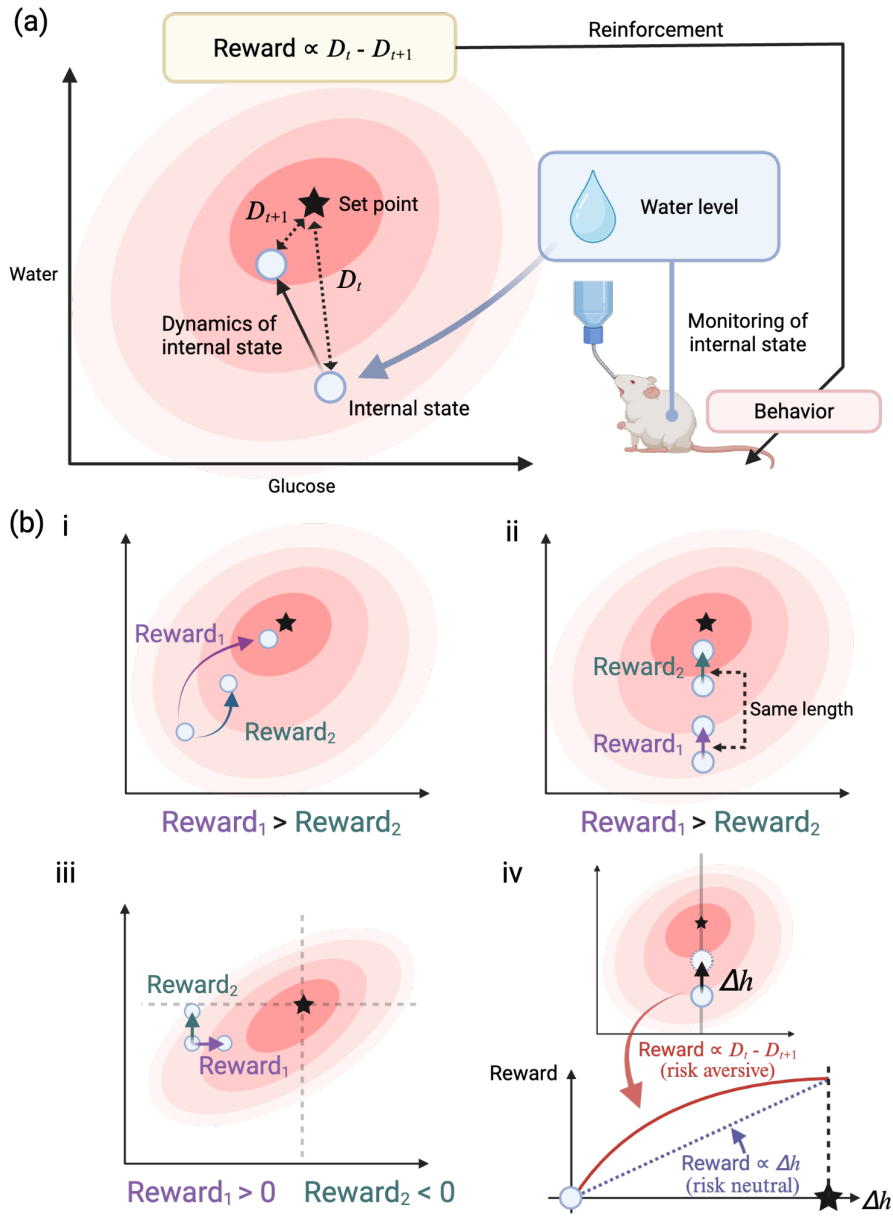

Figure 1. (a) Overview of HRRL. The agent monitors its internal bodily state (e.g., hydration level in the figure). Through interactions with the external environment, the internal state changes, moving closer to or farther from a set point. Rewards are determined based on changes in the distance from the set point, reinforcing the agent's behavior. (b) Properties of homeostatic rewards based on the characteristics of drive functions (Keramati and Gutkin 2011). (i) Reward value increases with the dose of outcome. (ii) Excitatory effect of deprivation level. (iii) Inhibitory effect of the irrelevant drive, under certain drive function. (iv) Risk aversion.

**HRRL and control theory**

HRRL is a normative theory - it produces physiologically rational agents. It can be mathematically shown that HRRL agents learn behaviors that minimize deviations from the homeostatic optimum in the long-term. More formally, HRRL policies minimize the sum of discounted future deviations or, equivalently, maximize future discounted homeostatically-defined rewards. Hence HRRL policies can be seen as optimal forward control strategies. We note that this demonstration of optimality requires discounting (see Keramati and Gutkin [8]).

Most earlier studies have considered HRRL not as a control problem but as a model-free RL problem (Keramati and Gutkin [9]): The agent learns appropriate actions in response to its internal state and its external state in the environment, but it does not have access to explicit models of i) the dynamics of its internal states, ii) the dynamics of the environment, and iii) the resource distribution in the environment. All these factors have to be inferred from data and implicitly shape the policy. Instead, if everything is known, the HRRL problem setting becomes an optimal control problem with an interesting bipartite structure. The goal is to minimize a cost function – the drive function – that depends only on the internal state. However, the agent's actions operate primarily on the external state and can only indirectly exert homeostatic control on the internal state by navigating the agent to external states with appropriate resources.

This control-theoretic formulation is beneficial for future work for a variety of reasons. Conceptually, it opens up a different perspective, in which biological agents are not foraging for rewards, but for control forces that regulate their internal state. Algorithmically, it allows the use of other optimization methods, depending on what is known about the agent and the environment. If everything is known, the host of methods from optimal control become available. This would, for example, provide an inroad to studying the complex and time-varying exploration-exploitation trade-offs inherent to HRRL, as discussed below. HRRL also opens up a new, intermediate form of "partially model-based" problems in RL, in which the agent knows the dynamics and the preference of its internal state – we can all predict when we will be hungry – but nothing about the external state dynamics and the resource distribution.

**HRRL accounts for a rich set of behaviors**

With a few assumptions on the drive function, HRRL yields a host of properties that chime in well with observed behavioral data (Figure 1b). Under a relatively generic parameter setting of the drive function (see Math Box), we can show that the associated value function also deviates from linearity in a way compatible with prospect theory as proposed by Kahneman and Tversky [13] - meaning it is sublinear. The resulting agents (1) are risk-averse and (2) loss-averse. Furthermore, values associated with outcomes are state-dependent: increasing deviation in a given direction potentiates the value, while deviating in a direction that is unrelated to this outcome, depresses its value. This cross-need competition indeed has been observed in behavior and successfully modeled within the HRRL framework (e.g., Uchida et al. [14], Ozawa et al. [15]). The state dependence of values also gives a normative non-linear model for incentive sensitization (augmenting the previously proposed ad hoc models (Berridge [16])).

When an internal state deviates from an appropriate condition—such as experiencing hunger—returning to a proper state (e.g., satiety) cannot typically be achieved instantaneously and internally; rather, it requires acquiring resources. Consequently, movement is necessary to obtain food, and as a side effect of controlling internal states, whole-body motion is autonomously acquired. This phenomenon has been observed in simulations where agent behavior is optimized using HRRL (Yoshida et al. [17]).

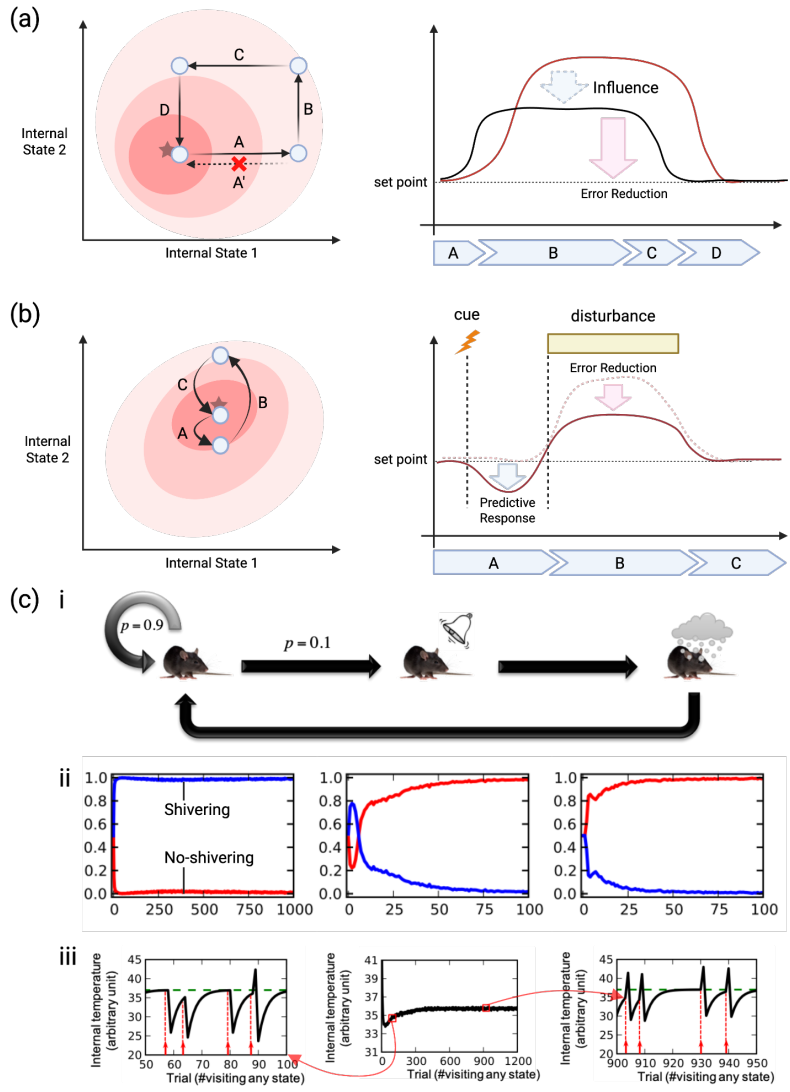

Figure 2. Behavioral properties of HRRL. (a) An example in which constraints on the dynamics of internal states result in a "detour" before minimizing the error. This panel considers a situation where an agent transitions from an optimal internal state to a different state due to some external factor and wants to return to its original state. In this case, due to constraints imposed by the body's internal systems, such as metabolism and the immune system, as well as external dynamics, it can be necessary to take the path A→B→C→D rather than the direct path A'. The panel on the right shows the corresponding changes in the internal state. An example of this is when immune signals are triggered by the invasion of viruses or bacteria, leading to a biological defense response and an increase in body temperature. The rise in body temperature affects the biological defense mechanisms within the body, contributing to the restoration of immune signals to their normal state. This increases short-term errors but reduces long-term cumulative errors. (b) An example of predictive error minimization when future disturbances are anticipated based on external signals. This panel shows that, under optimal internal states, when a cue stimulus indicating future disturbances is given, as shown in the panel on the right, a preliminary response A can be made to offset the disturbance when it actually occurs at B, thereby reducing the error. Such an anticipatory response is effective for minimizing the overall error when the error at A is less than the reduction in error at B. (c) An example of predictive shivering simulation. Here the simulated animal is exposed to a bell (CS) that predicts the subsequent thermal challenge. (i) state diagram of the simulated pavlovian thermal challenge task. (ii) Choice probability of shivering versus non-shivering action in (left panel) the initial state, (middle panel) the bell-ring state and (right panel) the thermal challenge state. Note that non-shivering is chosen consistently in the null state and the agent learns to predictively choose the shivering action in the bell-ring state. (iii) Panels show the internal state dynamics: the middle panel is the long-term internal state. We see that the agent learns to stabilize it at the optimal value (approx 35,5). Left panel and right panel show trial-by-trial internal state dynamics. Before learning (left panel) the internal state is consistently negatively deviated. Agent begins to learn to produce an anticipatory increase in internal temperature at the CS (marked by red arrow and dashed line). After learning the agent consistently produces an anticipatory response at the CS on each trial. The amplitude of this response is such that it optimally counters the ensuing thermal deviation (see Keramati and Gutkin 2011 for mathematical proof). Figure modified from Keramati and Gutkin (2014). For simulation details see Keramati and Gutkin 2014.

From the perspective of behavior as a byproduct of maintaining homeostasis, more complex behaviors are expected when the internal state variables are interdependent rather than independent. For example, when considering the dynamics of two internal states, it is possible that optimizing one state may temporarily increase the error in the other (Figure 2a). This phenomenon arises because HRRL makes predictive behavioral choices based on future states. If minimizing the cumulative error over a long-term timespan is possible, HRRL's optimal control may tolerate short-term increases in error. As a result, counterintuitive behaviors may emerge in the short term, depending on the interdependencies of internal bodily dynamics such as metabolic processes and the immune system.

Due to HRRL's ability to optimize over future time series, if an agent is provided with predictive cues about future disturbance in internal states, it can compensate in advance, further reducing long-term error (Figure 2b). This perspective naturally aligns with the concept of allostasis, which describes advanced regulation of bodily states through predictive control over extended timescales (Keramati and Gutkin [9]).

In the sense that rewards are computed internally, this approach is related to the concept of intrinsic motivation in artificial intelligence and machine learning (Oudeyer et al. [18]; Oudeyer and Kaplan [19]; Baldassarre [20]). Intrinsic motivation assigns internal rewards to unexplored states and prediction errors, thereby facilitating pretraining and exploration. Yet, the internal definition of rewards in HRRL and intrinsic motivation are not mutually exclusive but rather complementary; intrinsic motivation can be leveraged to enhance exploration in HRRL.

HRRL aims to satisfy multiple needs simultaneously, making exploration-exploitation trade-offs more complex than in classical RL. Agents must balance using known resources to maintain internal stability with exploring for new ones. For example, after finding water, an agent must exploit it for thirst while exploring for food. Model-based RL methods (Kearns and Singh [21]; Brafman and Tenenholtz [22]) offer a starting point for addressing this, though reward state-dependence would likely require modifications.

**Deep HRRL and artificial intelligence**

From an engineering perspective, the HRRL framework enables autonomous behavioral learning that meets multiple needs, without relying on an externally defined reward system outside the agent's body. Moreover, neuroscientists have discussed hypothetical robotic systems and their potential (Seth [23]; Solms [24]; Damasio [25]). As such, HRRL has recently attracted attention as a key component in developing biologically inspired artificial intelligence (Man and Damasio [26]; Moerland et al. [27]; Yoshida et al. [17]; Lee et al. [28]).

The scalability of HRRL has been extended into various problem domains, based on large-scale and highly nonlinear behavioral optimization (Mnih et al. [29]; Silver et al. [30]; Schulman et al. [31]). This includes systems with continuous vector observations and discrete action spaces (Yoshida [32]), continuous vector observations and continuous vector actions (Yoshida et al. [17]), image-based input with discrete action spaces (Yoshida [32]), high-dimensional multimodal input including images with continuous vector actions (Yoshida et al. [17]), and continuous-time environments (Laurencon et al. [33]). Furthermore, Dulberg has proposed a biologically inspired approach to decomposing reward functions in HRRL based on distinct motivational drives, suggesting its effectiveness (Dulberg et al. [34]). Intriguingly, the latter linked the multiple-drive version of HRRL to the notion of multiple selves in theories of consciousness. Moreover, HRRL has been shown to be a robust motivational framework capable of operating in real-world systems (Yoshida et al. [35]).

**HRRL and psychiatric disorders**

Recent works have also used HRRL to address multiple aspects of oriented nutrient behaviors and various behavioral pathologies. To address data on consummatory behaviors of nutrients, HRRL has been augmented to incorporate orosensory signals (see BOX and Duriez et al. [36]). Here the major conceptual proposal is that the gustatory signal associated with a nutritive resource (e.g., food items, sodium, etc.) is an unbiased predictor of that resource's impact on the internal state. In turn, this orosensory information can be either innate and/or learned (e.g., by an error-correction process).

Hall and colleagues have made a recent proposal that certain aspects of consummatory depression can be modeled as a pathology of the drive function (Hall et al. [37]), while an allostatically extended version of the HRRL was proposed as a model for cocaine addiction (Keramati et al. [38]). In the latter, the authors linked both positive and negative motivational theories of addiction and showed how such can lead to emergent drug-intake escalation, drug-dose dependent relapses. This opens the horizons to start thinking about psychiatric disorders as resulting from pathologies of interoceptive motivation (Petzschner et al. [39]; Keramati et al. [38]; Yee [40]).

**Conclusion**

In this perspective, we have focused on basic physiological needs and drives. For these, there is ample evidence that neural afferents tracking physiological internal variables impinge on the dopaminergic structures that signal reward and reward predictions, such as the VTA (see Weber et al. [41]). These dopaminergic signals are in turn modulated by changes in the internal state of the animal in a manner compatible with the HRRL theory (e.g., see Roitman et al. [42]; Hsu et al. [43]; Grove et al. [44]). Yet, several proposals have been made that the HRRL framework can be applied beyond such basic drives. For example, Juechems et al. suggested that monetary values may be formed by exactly the same process, with humans setting internal financial goals that serve as de facto set points (Juechems and Summerfield [45]). In fact, HRRL may also be applied to social phenomena. Already Yoshida et al. (Yoshida and Man [46]) explored cross-agent tracking of internal drives may lead to emergent prosocial behaviors. Intriguingly, recent evidence suggests that signals tracking the social state of the animal may be signaled by hypothalamic circuits to the dopaminergic structures in the same "homeostatic" manner as basic physiological ones (Liu et al. [47]) and that dopaminergic system indeed tracks social rewards in a manner analogous to primary physiological ones (e.g., Solié et al. [48]). An intriguing challenge to the HRRL theory is the recent evidence that social, preservation, and even parenting behaviors are modulated by hunger states (Alcantara et al. [49]; de Araujo Salgado et al. [50]; Pozo et al. [51]) through shared neural circuits (Isaac and Murugan [52]) - we venture to speculate that our theory is well positioned to understand such interactions through the prism of drive-based multi-objective learning and control.

We believe that the HRRL is a computational framework that allows for ample avenues to explore the origins of reward and the role of interoception in structuring motivated behaviors. On the other hand, HRRL may be a new direction for autonomous and robust learning in artificial systems.

**Acknowledgements**

This research is supported by Japan Society for the Promotion of Science KAKENHI grant 24K23892, BSG is funded Agence Nationale pour la Recherche (ANR-17-EURE-0017, CogFinAgent), ENS, CNRS, and INSERM.

**Declaration of generative AI and AI-assisted technologies in the writing process.**

During the preparation of this work, the author(s) used BioRender.com in order to create illustrative figure diagrams. ChatGPT (4o, services that maintain the confidentiality of information) was used by NY to revise English grammar. After using this tool/service, the author(s) reviewed and edited the content as needed and take(s) full responsibility for the content of the publication.

**References**

Papers of particular interest, published within the period of review, have been highlighted as: * of special interest; ** of outstanding interest.

Comment: Authors extend the homeostatically regulated RL framework to consider multiple independent homeostatic sub-agents, each learning an independent need and that compete at the action selection stage. Results suggest such modular HRRL architecture improves efficiency in complex environments.

Comment: In this perspective Hall et al. link pathological patterns of behavior linked with depression-related anhedonia to distortions of the drive function. This proposes that changes in the internal state representations may lead to avoidance of otherwise rewarding experiences.

44. *Grove JCR, Gray LA, La Santa Medina N, et al. Dopamine subsystems that track internal states. Nature. 2022;608(7922):374-380.

Comment: The study identifies a hypothalamic pathway that signals fluid balance to the VTA, which then communicates with circuits monitoring ingestion. Using a paradigm to separate oral and post-absorptive effects, it shows mice can learn to prefer fluids based on rehydration properties. This may be key biological evidence for the interplay between the orosensory and internal impact of consumed sources as proposed in the HRRL.

45. Juechems K, Summerfield C. Where does value come from? Trends Cogn Sci. 2019;23(10):836-850.
46. Yoshida N, and Man L. "Empathic Coupling of Homeostatic States for Intrinsic Prosociality." Intrinsically-Motivated and Open-Ended Learning Workshop@ NeurIPS2024.
47. **Liu D, Rahman M, Johnson A, Amo R, Tsutsui-Kimura I, Sullivan ZA, et al. A hypothalamic circuit underlying the dynamic control of social homeostasis. Nature. 2025;640:1000–1010.

Comment: This identifies key specific neural pathways and mechanisms within the hypothalamus that regulate social behavior and interactions. These findings suggest that the brain actively maintains a stable balance in social engagement that may be learned through a homeostatically regulated reinforcement learning.

48. Solié C, Girard B, Righetti B, Tapparel M, Bellone C. VTA dopamine neuron activity encodes social interaction and promotes reinforcement learning through social prediction error. Nat Neurosci. 2022;25(1):86-97.
49. Alcantara IC, et al. A hypothalamic circuit that modulates feeding and parenting behaviors. bioRxiv. 2024.07.22.604437.
50. de Araujo Salgado I, et al. Toggling between food-seeking and self-preservation behaviors via hypothalamic response networks. Neuron. 2023;111(18):2899-2917.e6.

Comment: Study shows evidence for interactions between internal state stability and social interactions. These findings suggest that physiological internal state variables and social status internal states may indeed be modeled together in a homeostatic reinforcement learning framework.